\documentclass[10pt, conference, letterpaper, bottom=0.5in, top=0.7in]{IEEEtran}

\IEEEoverridecommandlockouts

\usepackage{lipsum}
\usepackage{hyperref}
\usepackage{scalerel}
\usepackage{tikz}
\usetikzlibrary{svg.path}
\usepackage{xcolor}
\usepackage{cite}
\usepackage{algorithm}
\usepackage{algpseudocode}
\usepackage{stackengine}
\usepackage{amsmath,amssymb,amsfonts}
\usepackage{graphicx}
\usepackage{textcomp}
\usepackage{xcolor}
\usepackage[utf8]{inputenc}
\usepackage{fixmath}
\usepackage{balance}
\usepackage{multirow}
\usepackage{xparse}
\NewDocumentCommand{\Log}{o}{%
  \IfNoValueTF{#1}{}{{}^{#1}\!}\log}%
\usepackage{array}
\usepackage{booktabs}
\usepackage{soul}
\usepackage{hyperref}
\usepackage{lineno}
\usepackage{inputenc}

\usepackage{booktabs}% http://ctan.org/pkg/booktabs
\newcommand{\tabitem}{~~\llap{\textbullet}~~}

\usepackage{subcaption}
\usepackage{authblk}
\usepackage[justification=centering]{caption}
\usepackage[export]{adjustbox}
\usepackage{mathtools}

\def\BibTeX{{\rm B\kern-.05em{\sc i\kern-.025em b}\kern-.08em
    T\kern-.1667em\lower.7ex\hbox{E}\kern-.125emX}}

\foreach \x in {A, ..., Z}{
	\expandafter\xdef\csname orcid\x\endcsname{\noexpand\href{https://orcid.org/\csname orcidauthor\x\endcsname}{\noexpand\orcidicon}}
}

\usepackage[inline]{./trackchanges}
\addeditor{Rizqi}

\begin{document}

\title{Guard Beam: Protecting mmWave Communication through In-Band Early Blockage Prediction}

\author[1]{Rizqi Hersyandika}
\author[1,2]{Yang Miao}
\author[1]{Sofie Pollin}

\affil[1]{Department of Electrical Engineering, KU Leuven, Belgium}
\affil[2]{Department of Electrical Engineering, University of Twente, the Netherlands}

\maketitle

\begin{abstract}

Human blockage is one of the main challenges for mmWave communication networks in dynamic environments. The shadowing by a human body results in significant received power degradation and could occur abruptly and frequently. A shadowing period of hundred milliseconds might interrupt the communication and cause significant data loss, considering the huge bandwidth utilized in mmWave communications. An even longer shadowing period might cause a long-duration link outage. Therefore, a blockage prediction mechanism has to be taken to detect the moving blocker within the vicinity of mmWave links. By detecting the potential blockage as early as possible, a user equipment can anticipate by establishing a new connection and performing beam training with an alternative base station before shadowing happens. This paper proposes an early moving blocker detection mechanism by leveraging an extra guard beam to protect the main communication beam. The guard beam is intended to sense the environment by expanding the field of view of a base station. The blockage can be detected early by observing received signal fluctuation resulting from the blocker's presence within the field of view. We derive a channel model for the pre-shadowing event, design a moving blockage detection algorithm for the guard beam, and evaluate the performance of the guard beam theoretically and experimentally based on the measurement campaign using our mmWave testbed. Our results demonstrate that the guard beam can extend the detection range and predict the blockage up to 360 ms before the shadowing occurs.

\end{abstract}


\IEEEpeerreviewmaketitle

\section{Introduction}
\label{sec_introduction}

Millimeter-Wave (mmWave) frequencies are favoured in 5G and beyond-5G wireless communication networks because of the large bandwidth available for high achievable throughput~\cite{Rappaport2013}. A mmWave link uses narrow directive beams at communicating stations to compensate for the high free-space propagation loss. However, the directive beams make the mmWave link easily obstructed or blocked by moving objects/humans, causing severe performance degradation.      

Human activities in dynamic environments, such as walking across the Fresnel zone of a mmWave link, could significantly affect the mmWave communication performance. Power attenuation caused by the human body, including the scattering\cite{9560206} and absorption, can reach up to 34~dB, depending on the frequency band\cite{Zhao2017}. The shadowing time caused by human walking across a mmWave link can last for several hundreds of milliseconds\cite{Slezak2018}. Therefore, it is necessary to predict the human blockage before it shadows the mmWave communication link and take the actions required to cope with it, such as re-steering the beam to find an alternative path or establishing a new connection with another base station. Joint communication and sensing are envisioned in 6G networks, where sensing the surrounding environment in the presence of humans is as important as achieving high data rate communication \cite{9330512}.

Reported methods for predicting the human blockage in mmWave communications include the use of 1) vision/camera\cite{Oguma2016,Koda2020,Charan2021}, 2) radar\cite{demirhan2021radar}, 
3) LiDAR\cite{Marasinghe2021}, and 4) sub-6 GHz channel\cite{Alrabeiah2020,Ziad2020}. 
However, those methods require and rely on an additional system to predict the moving blockage, thus adding cost and complexity to the implementation.

The predictive blockage-preventing approach using an in-band mmWave signal and data rate observation is proposed in \cite{Alkhateeb2021,Koda2020}. The pre-blockage radio frequency signature, the fluctuation of the received (Rx) signal strength level occurring before the shadowing event, is utilized to predict the future time instance of a blockage \cite{Alkhateeb2021}. It is shown that the moving blocker can be detected before shadowing occurs by leveraging the signal fluctuation observed by the mmWave communication beam. 
The prediction accuracy decreases as the blocker is far from the mmWave link, meaning that it can only be detected accurately when it nearly approaches the mmWave communication beam. In \cite{Koda2020}, the data rate fluctuation occurring
before the shadowing indicates the potential blockage. However, the achievable average prediction time is quite limited, i.e., up to 90~ms, leaving insufficient time for the countermeasure of the blockage (e.g., handover), which typically requires several hundreds of milliseconds. Therefore, an early moving blockage detection mechanism is essential for reliable mmWave communications.

\begin{figure}[t]
    \centering
    \includegraphics[width=0.95\linewidth]{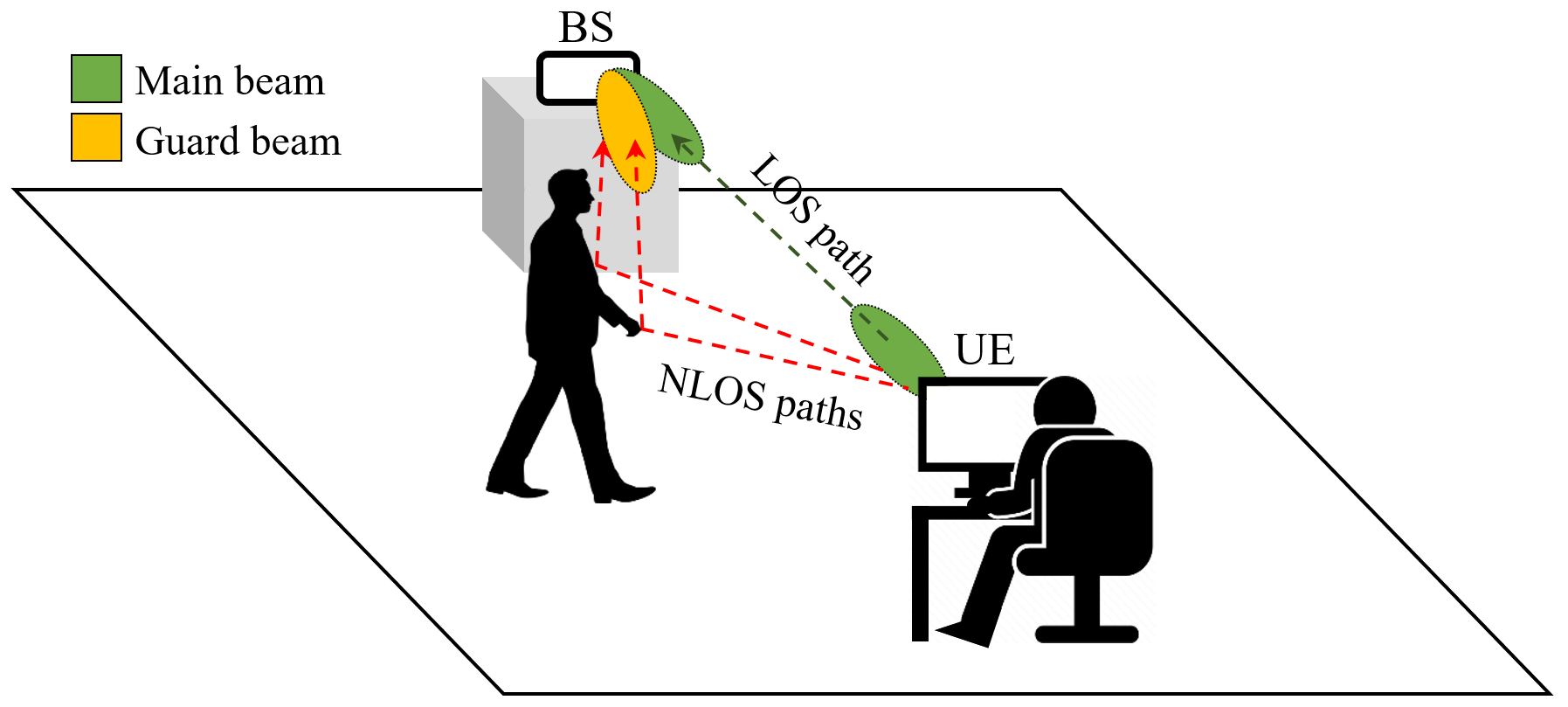}
    \caption{\centering{Illustration of blockage detection using a guard beam}} 
    \label{fig:illustration}
\end{figure}

Therefore, an early moving blockage detection mechanism is essential for reliable mmWave communications. This paper proposes a dynamic blockage prediction method using a mmWave guard beam, an additional passive mmWave Rx beam in addition to the main communication beam. 
As illustrated in Fig.~\ref{fig:illustration}, the guard beam aims to detect early the presence of a moving blocker before the main communication link suffers from shadowing. The proposed early detection is achieved by observing the Rx signal level fluctuation resulting from the non-line-of-sight (NLOS) component from the User Equipment (UE) to the guard beam of the Base Station (BS). The existence of the guard beam extends the field of view of BS, thus giving sufficient time for BS and UE to take action in the prevention of blockage. Moreover, by leveraging an extra mmWave Rx beam, no additional system but only an extra RF chain is required on BS for the blockage prediction.  

Contributions of this paper are summarized as follows:
\begin{enumerate}
    \item We propose an early dynamic blockage prediction mechanism by leveraging an extra passive guard beam implemented as a receiver in BS. The guard beam extends the BS field of view, senses the environment, and detects the potential moving blockers early.
    
    \item We derive the channel models of the pre-shadowing event for both the main communicating beam and the guard beam. This blocker-position-dependent model can detect the presence of a moving blocker approaching the communication beam. 

   \item We analyze the detection capability of different configurations of the main and guard beams and evaluate the guard beam's performance theoretically and experimentally based on a measurement campaign using our mmWave testbed. 
   It is found that the guard beam can extend the detection range up to 860 mm and predict the blockage up to 360 ms before the shadowing occurs.
    
\end{enumerate}

In Sec.~\ref{sec:models}, we derive the mmWave channel models for both the main communication beam and the sensing guard beam for the pre-shadowing event. The blockage detection algorithm is also explained. In Sec.~\ref{sec:analysis}, we analyze the detection capability of various configurations of main and guard beams. In Sec.~\ref{sec:measurements}, we evaluate the prediction time of the proposed method based on measured data. Finally, we conclude the paper in Sec.~\ref{sec:conclusion}.

\section {System Model and Algorithm Design}
\label{sec:models}

\subsection{Pre-shadowing channel model}

\begin{figure}[t]
  \subfloat[Main beam only]{
	\begin{minipage}[b]{0.165\textwidth} \label{fig:fov_main_beam}
	   \centering
	   \includegraphics[width=\textwidth]{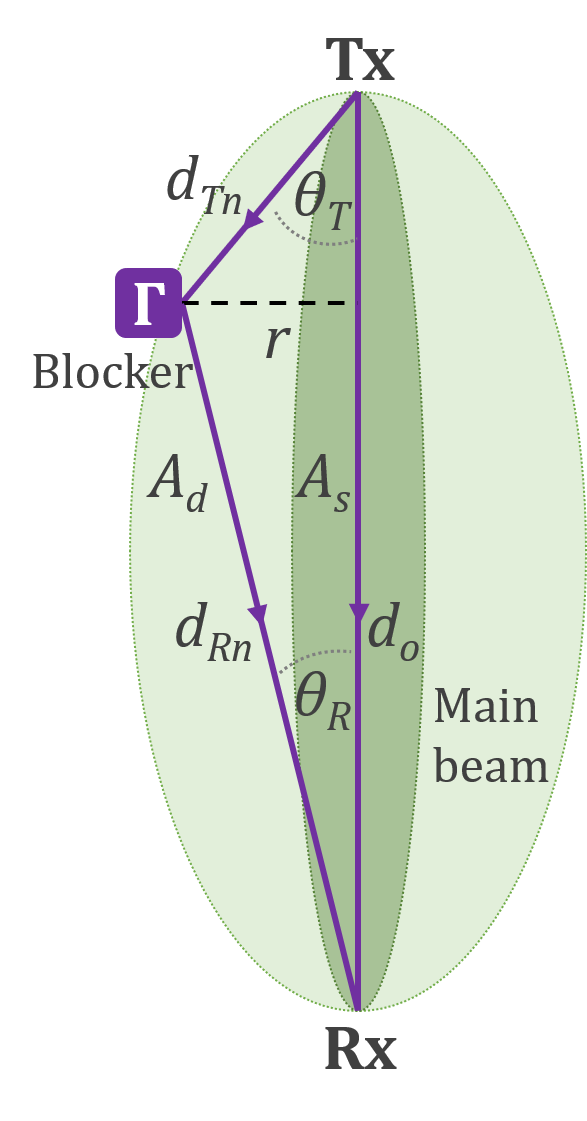}
	\end{minipage}}%
  \subfloat[Main beam \& guard beam]{
	\begin{minipage}[b]{0.257\textwidth} \label{fig:fov_guard_beam}
	   \centering
	   \includegraphics[width=\textwidth]{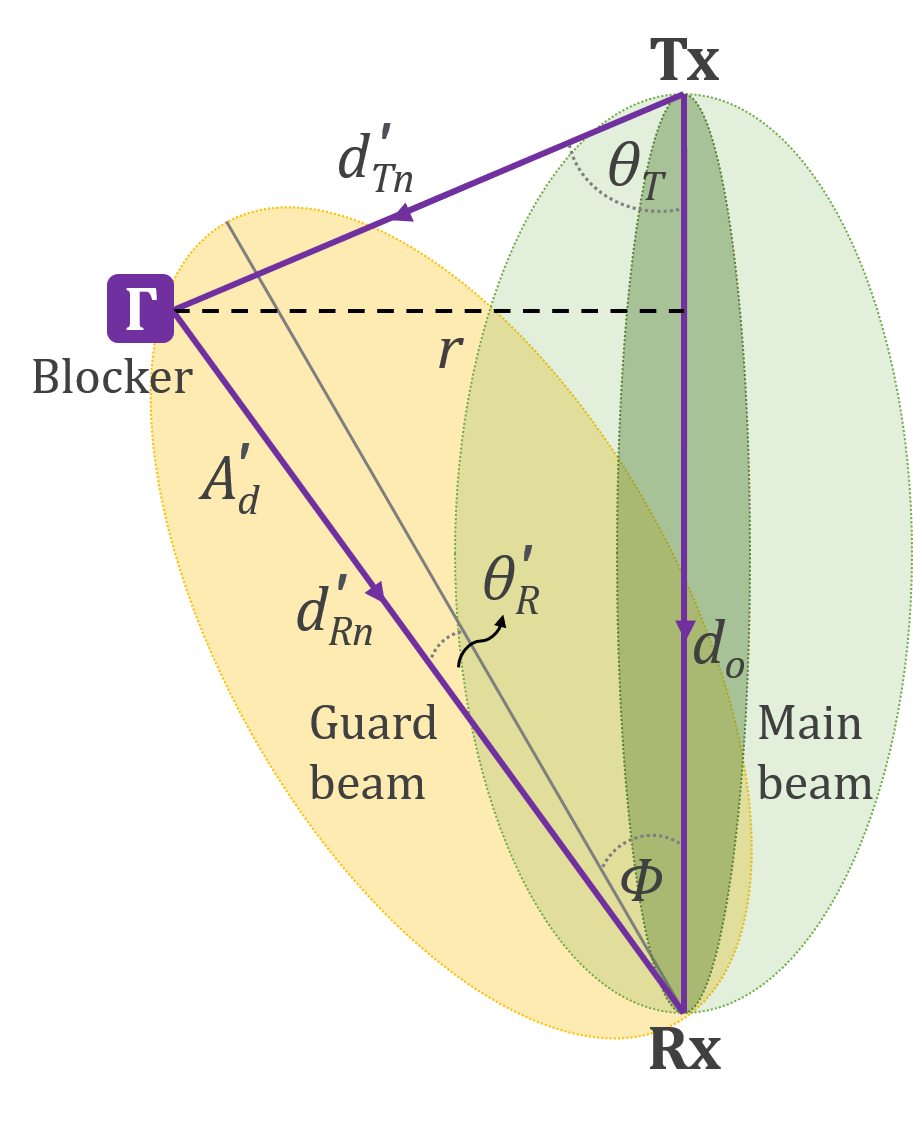}
	\end{minipage}}
\caption{Rx's 2-D geometrical field of view}
\end{figure}

During the pre-shadowing event, the movement of a blocker towards the main communication beam can be detected when the blocker is within the field of view of the Rx beam, as indicated by the outer green ellipse in Fig.~\ref{fig:fov_main_beam}. By having an additional guard beam whose focusing direction is in $\Phi$ angle separation from the main beam's focusing direction, the Rx's field of view is expanded with regard to the beam width and separation angle, as demonstrated by the yellow ellipse in Fig.~\ref{fig:fov_guard_beam} in addition to the green ellipse. {Consequently, the detection range $r$, defined as the maximum detectable distance of the blocker to the direct path of the communication link, increases.} 
The blockage detection area is modelled in a 2-D plane, which is transferable to a 3-D scenario.

We consider $\hat{x}$ as the transmitted signal from the transmitter (Tx), then the received signal at Rx's main beam becomes 
\begin{equation}
    \hat{y}(p) = \sqrt{p_T}\hat{h}(p)\hat{x} + \hat{w},
\end{equation}
and the received signal at Rx's guard beam is formulated as
\begin{equation}
    \hat{y}'(p) = \sqrt{p_T}\hat{h}'(p)\hat{x} + \hat{w},
\end{equation}
where $\hat{h}(p)$ and $\hat{h}'(p)$ represent the channel for the main and guard beam, respectively, and depend on the blocker's position $p$, $p_T$ is the transmit power and $\hat{w}$ is the additive noise at the receiver. The blocker's position relative to Tx and Rx is defined as $p(r, \theta_{T}, \theta_{R})$, which depends on the detection range $r$, the azimuth angle between Tx and blocker $\theta_{T}$ and the azimuth angle between Rx's main beam and blocker $\theta_{R}$. By defining the steering angle differences between the main and guard beam as $\Phi$, the azimuth angle between Rx's guard beam and blocker becomes $\theta_{R}' = \theta_{R} - \Phi$.

\subsubsection{Main beam}

The main communication beam detects the moving blocker when the blocker position $p$ is within the main beam's detection area, as illustrated in Fig.~\ref{fig:fov_main_beam}. During the pre-shadowing event, the blocker-position-dependent channel of the main communication beam pair is defined as:
\begin{equation}
    \hat{h}(p)=\begin{cases}
        \hat{h_{o}}, & p \notin A_{d}\\
        \hat{h_{o}} + \sum\limits_{n=1}^{N} \hat{h_{n}}(p), & p \in A_{d}, p \notin A_{s}.\\
  \end{cases}
\end{equation}
where the detection area $A_d$ is the area where the influence of the blocker is as the interference to the Line-Of-Sight (LOS) path, and the shadowing area $A_s$ is the area where the influence of the blocker is to obstruct the LOS path. 
When the blocker is outside $A_d$, a LOS component $\hat{h_{o}}$ between Tx and Rx dominates the channel. When the blocker enters $A_d$ from outside of it, some signals transmitted in the direction of the blocker are reflected off the blocker to the Rx main beam resulting in additional $N$ NLOS multipath components $\hat{h_{n}}(p)$. These NLOS components highly depend on the blocker's position. 
This model is only valid when the blocker is outside of $A_s$. Otherwise, the shadowing and diffraction would affect the direct path. Nevertheless, the channel characteristic during the shadowing event is beyond the scope of this paper as we aim at predicting moving blockage before shadowing occurs. 

The main beam's LOS component is formulated as follows:
\begin{equation}
    \hat{h_{o}} = g_{T}(\Theta_{T},\theta_{T_{o}}) g_{R}(\Theta_{R},\theta_{R_{o}}) \frac {\lambda}{4 \pi d_{o}} e^{-j \frac { 2 \pi} {\lambda} d_{o}},
\end{equation}
where $g_T$ and $g_R$ represent the Tx and Rx array gain depending on the Half Power Beam Width (HPBW) of Tx $\Theta_{T}$ and Rx $\Theta_{R}$, as well as the azimuth LOS direction of Tx $\theta_{T_{o}}$ and Rx $\theta_{R_{o}}$, respectively. Typically, these Tx and Rx gains in this direction are maximal, thanks to the prior beam training. $\lambda$ is the wavelength, and $d_{o}$ is the LOS distance between Tx and Rx.
On the other hand, the blocker-position-dependent NLOS component is expressed as follows: 

\begin{equation}
    \hat{h_{n}}(p) = g_{T}(\Theta_{T},\theta_{T_{n}}) g_{R_{n}}(\Theta_{R},\theta_{R_{n}}) \Gamma_{n} \frac {\lambda}{4 \pi d_{n}(p)} e^{-j \frac { 2 \pi} {\lambda} d_{n}(p)},
\end{equation}
where the transmit and receive gain depend on the transmit angle $\theta_{T_{n}}$ and receive angle $\theta_{R_{n}}$, respectively. $\Gamma_{n}$ is the reflection coefficient of the $n$-th NLOS component. The total NLOS distance $d_{n} = d_{T_n} + d_{R_n}$, and thus:
\begin{equation}
    d_{n} = \frac{r}{sin({\theta_{T_{n}}})} + \frac{r} {sin({\theta_{R_{n}}})}.
\end{equation}

\begin{figure}[tb]
  \subfloat[Detection window]{
	\begin{minipage}[b]{0.244\textwidth} \label{fig:sliding_window}
	   \centering
	   \includegraphics[width=1.0\textwidth]{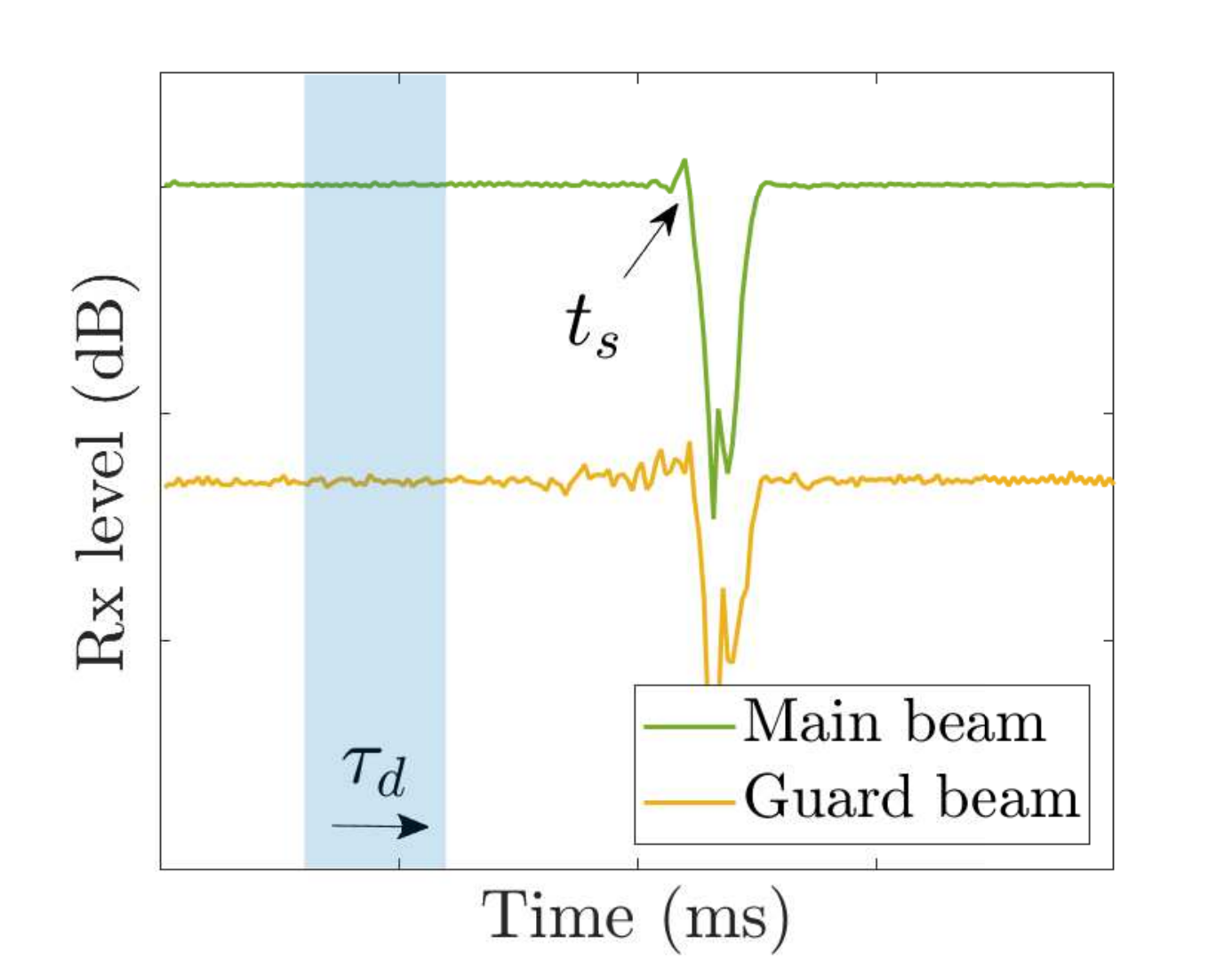}
	\end{minipage}}%
	\subfloat[Standard deviation threshold]{
	\begin{minipage}[b]{0.244\textwidth} \label{fig:std_threshold}
	   \centering
	   \includegraphics[width=\textwidth]{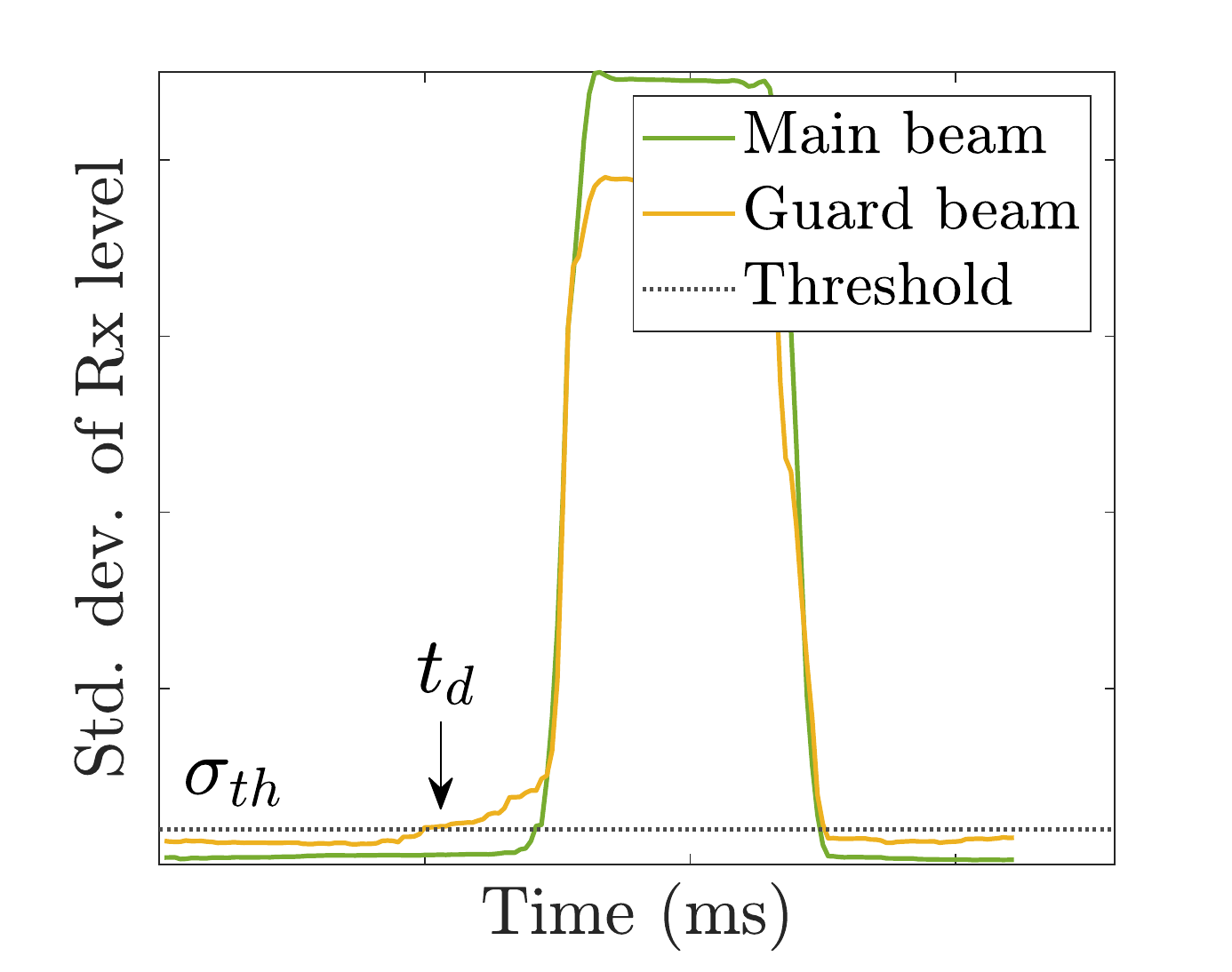}
	\end{minipage}}
\caption{Blockage detection parameters}
\end{figure}
\vspace{-1mm}

\subsubsection{Guard beam}
As described in Fig.~\ref{fig:fov_guard_beam}, the detection area of a guard beam is defined as $A_{d}'$. During the pre-shadowing event, the guard beam channel is defined as follows:
\begin{equation}
    \hat{h}'(p)=\begin{cases}
        \hat{h_{o}}', & p \notin A_{d}'\\
        \hat{h_{o}}' + \sum\limits_{n=1}^{N} \hat{h_{n}}'(p), & p \in A_{d}',\\
  \end{cases}
\end{equation}
where only a LOS component $\hat{h_{o}}'$ exists when the blocker is outside of $A_{d}'$, and additional NLOS components $\hat{h_{n}}'(p)$ are added when the blocker is inside of $A_{d}'$.
The LOS component of a guard beam is formulated as follows:
\begin{equation}
    \hat{h_{o}}' = g_{T}(\Theta_{T},\theta_{T_{o}}) g_{R}'(\Theta_{R}',\theta_{R_{o}}') \frac {\lambda}{4 \pi d_{o}} e^{-j \frac { 2 \pi} {\lambda} d_{o}}
\end{equation}
where $g_{R}'$ is the Rx gain of the guard beam depending on the angle of arrival concerning the guard beam's direction $\theta_{R_{o}}'$. The NLOS component of a guard beam is expressed as follows:
\begin{equation}
\begin{split}
    \hat{h_{n}}'(p) = g_{T_{n}}(\Theta_{T},\theta_{T_{n}}) g_{R_{n}}'(\Theta_{R}', \phi, \theta_{R_{n}}') \Gamma_{n} \\
    \frac {\lambda}{4 \pi d_{n}'(p)} e^{-j \frac { 2 \pi} {\lambda} d_{n}'(p)}.
\end{split}
\end{equation}
In this case, the separation angle between the main and guard beam $\Phi$ determines the $g_{R_{n}}'$. The NLOS distance becomes $d_{n}' = d_{T_n}' + d_{R_n}'$, and thus: 
\begin{equation}
    d_{n}' = \frac{r}{sin({\theta_{T_{n}}})} + \frac{r} {sin({\phi + \theta_{R_{n}}'})}.
\end{equation}

\begin{algorithm}
\caption{Blockage detection algorithm}\label{alg:cap}
\begin{algorithmic}

\Require $\hat{y}, \hat{y}', \tau_{d}, \sigma_{th}$ 
\State $ z_{t} = |\hat{y_{t}} + \hat{y_{t}}'| $
\State $\sigma(t) = std\{z_{t},z_{t-1},...,z_{t-\tau_{d}}\}$ 

\While{$\sigma(t) < \sigma_{th}$}
    \Comment{Blockage is not detected}
    \State{$t = t+1$}
    \State $\sigma(t) = std\{z_{t},z_{t-1},...,z_{t-\tau_{d}}\}$ 
\EndWhile   
\Comment{Blockage is detected}

\State $t_d \gets t$
\Comment{Detection time}

\end{algorithmic}
\label{detection_algorithm}
\end{algorithm}

\subsection{Blockage detection algorithm}

The blocker's presence is indicated by the received signal level fluctuation when it enters the prediction area. This received signal level volatility increases the standard deviation of the received signal level over an observation period.
We define the observation period as the detection window $\tau_d$, as shown in Fig.~\ref{fig:sliding_window}. The time instance when the Rx level of the main beam starts decreasing due to shadowing is called the shadowing time $t_s$. 

The blockage is detected when the standard deviation of received signal level $\sigma(t)$ over the detection window exceeds a threshold $\sigma_{th}$. The time instance when $\sigma(t) \geq \sigma_{th}$ is defined as the blockage detection time $t_d$, as indicated in Fig.~\ref{fig:std_threshold}. The difference between $t_s$ and $t_d$ is defined as the prediction time $t_p$, which is the main metric to evaluate. Algorithm~\ref{detection_algorithm} explains the blockage detection algorithm.
Different $\sigma_{th}$ value needs to be assigned for different Rx beam configuration due to various level of typical standard deviation values.

\subsection{Guard beam implementation}
A guard beam is a passive beam next to the main communication beam that only has a receiving capability. Considering the hardware limitation of a UE, the guard beam is preferable to be implemented on the BS side. A BS can take advantage of the uplink training pilot transmitted by a UE to scan and detect the potential blockage using a guard beam in which the scanning period depends on $\tau_d$. 

The implementation of a guard beam requires an extra dedicated RF chain and several antenna elements to generate an independent Rx beam with a beamwidth of $\Theta_{R}'$. A single guard beam can be swept to the left and right sides of the main beam alternately to protect the main beam in two azimuth directions. Alternatively, two-sided guard beams can be implemented at the cost of two additional RF chains other than the main communication beam RF chain.

\section{Numerical Analysis}
\label{sec:analysis}

\begin{table}[b]
\vspace{0.25cm}
\centering
\setlength\extrarowheight{1pt}
\vspace{-5mm}
\caption{Simulation and measurement parameters}
\label{tab:parameters}
\captionsetup{size=footnotesize,
    skip=2pt, position = bottom}
{
\begin{tabular}{|l|c|c|}
\hline
\multicolumn{1}{|c|}{\textbf{Parameter}} & \multicolumn{1}{c|}{\textbf{Simulation}} & \multicolumn{1}{c|}{\textbf{Measurement}} \\
\hline\hline
Frequency ($f$) & \multicolumn{2}{c|}{26.0 GHz} \\ \hline  
Transmit power ($p_T$) & \multicolumn{2}{c|}{1 mW} \\ \hline
Tx HPBW ($\Theta_{T}$) & \multicolumn{2}{c|}{7°} \\ \hline
Rx HPBW ($\Theta_{R}$, $\Theta_{R}'$) & 7° and 13° & 7° \\ \hline
Steering angle ($\Phi$) & \multicolumn{2}{c|}{7° and 14°} \\ \hline 
Reflection coeff. ($\Gamma$) & 0.62 \cite{owda2020reflectance} & - \\ \hline 
Avg. noise (${\bar{w}}$) & -93.8 dBm & -100.8 dBm\\ \hline
Additional Rx gain  & - & 31 dB\\ \hline
Tx-Rx distance ($d_o$) & \multicolumn{2}{c|}{5 m} \\ \hline
Tx and Rx height & - & 1 m \\ \hline
Detection window size ($\tau_{d}$) & \multicolumn{2}{c|}{100 ms} \\ \hline
Std. dev. threshold ($\sigma_{th}$): &  &  \\ 
\tabitem Main only ($\Theta_{R}$ = 7°) & 0.03 & 0.035 \\
\tabitem Main only ($\Theta_{R}$ = 13°) & 0.1 & - \\ 
\tabitem Guard ($\Theta_{R}'$ = 7°, $\Phi$ = 7°) & 0.03 & 0.13 \\  
\tabitem Guard ($\Theta_{R}'$ = 13°, $\Phi$ = 7°) & 0.03 & - \\ 
\tabitem Guard ($\Theta_{R}'$ = 7°, $\Phi$ = 14°) & 0.03 & 0.58 \\ 
\tabitem Guard ($\Theta_{R}'$ = 13°, $\Phi$ = 14°) & 0.03 & - \\ 
\hline
\end{tabular}}
\end{table}

The pre-shadowing channel model for the main and guard beams is evaluated through the simulation in a 2-D plane to estimate the Rx field of view and detection range $r$. We consider a Uniform Linear Array (ULA) with a different number of elements to generate various beamwidth configurations for $\Theta_R$ and $\Theta_R'$, and apply different steering angles $\Phi$ to the guard beam. We evaluate $r$ for a fixed Tx-Rx distance $d_o$ using a single detection window $\tau_d$. Different standard deviation threshold $\sigma_{th}$ is assigned to each beam configuration due to the different levels of Rx fluctuation. The complete simulation parameters are listed in Table~\ref{tab:parameters}.

\begin{figure}[tbh]
  \subfloat[Main only: $\Theta_{R}$=7\textdegree]{
	\begin{minipage}[b]{0.24\textwidth} \label{fig:fov_main_7}
	   \centering
	   \includegraphics[width=\textwidth]{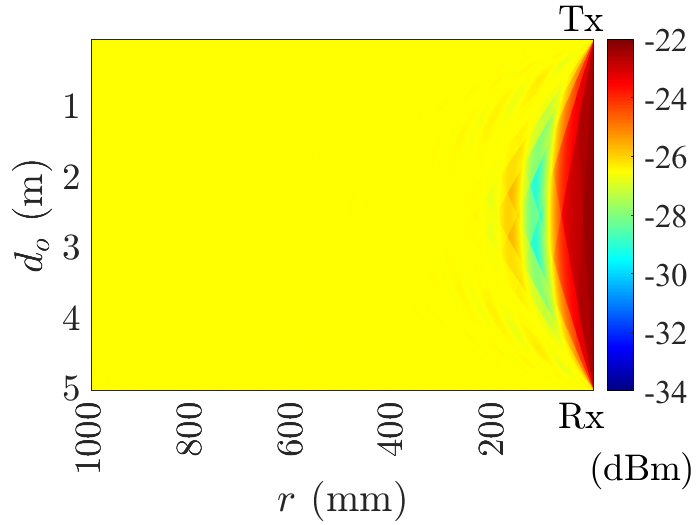}
	\end{minipage}}%
	\subfloat[Main only: $\Theta_{R}$=13\textdegree]{
	\begin{minipage}[b]{0.24\textwidth} \label{fig:fov_main_13}
	   \centering
	   \includegraphics[width=\textwidth]{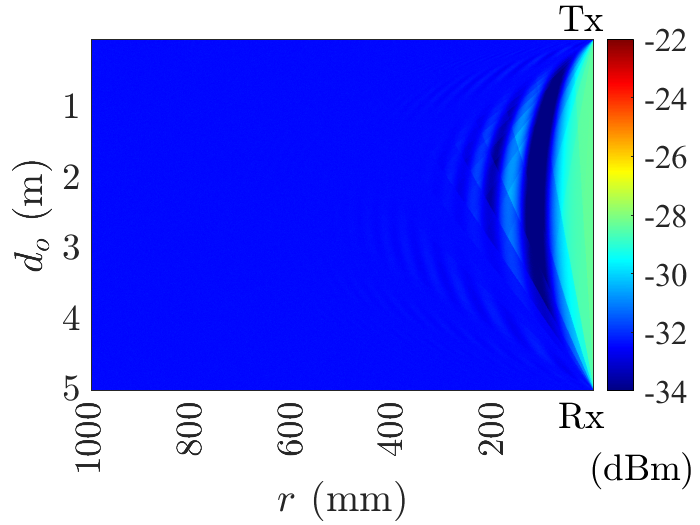}
	\end{minipage}}
	\newline
  \subfloat[Guard: $\Theta_{R}'$=7\textdegree, $\Phi$=7\textdegree]{
	\begin{minipage}[b]{0.24\textwidth} \label{fig:fov_guard_7_7}
	   \centering
	   \includegraphics[width=\textwidth]{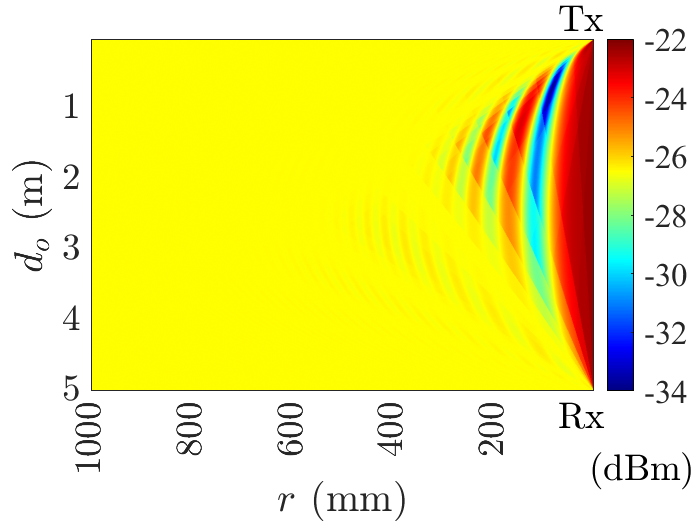}
	\end{minipage}}%
	\subfloat[Guard: $\Theta_{R}'$=13\textdegree, $\Phi$=7\textdegree]{
	\begin{minipage}[b]{0.24\textwidth}
	   \centering
	   \includegraphics[width=\textwidth]{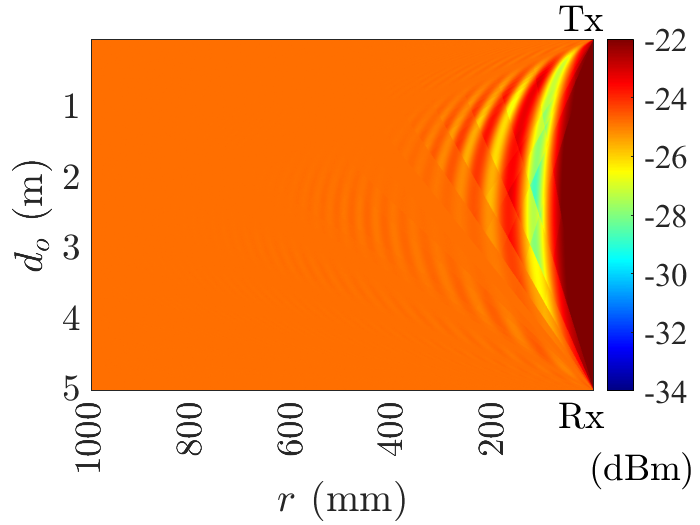}
	\end{minipage}}
	\newline
   \subfloat[Guard: $\Theta_{R}'$=7\textdegree, $\Phi$=14\textdegree]{
	\begin{minipage}[b]{0.24\textwidth} \label{fig:fov_guard_7_14}
	   \centering
	   \includegraphics[width=\textwidth]{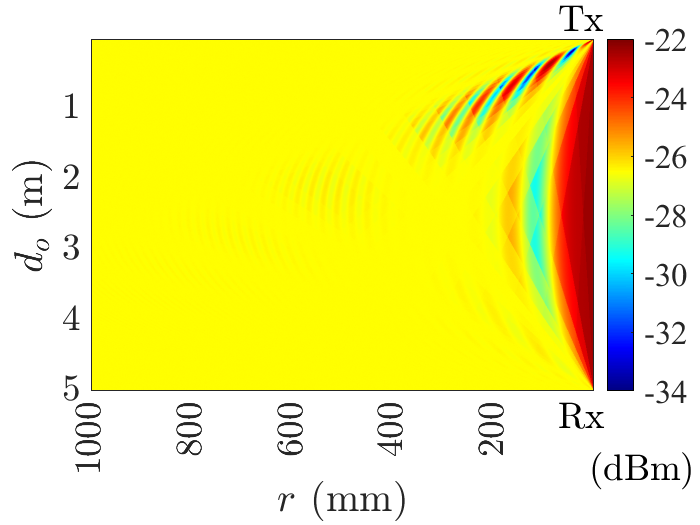}
	\end{minipage}}%
	\subfloat[Guard: $\Theta_{R}'$=13\textdegree, $\Phi$=14\textdegree]{
	\begin{minipage}[b]{0.24\textwidth}
	   \centering
	   \includegraphics[width=\textwidth]{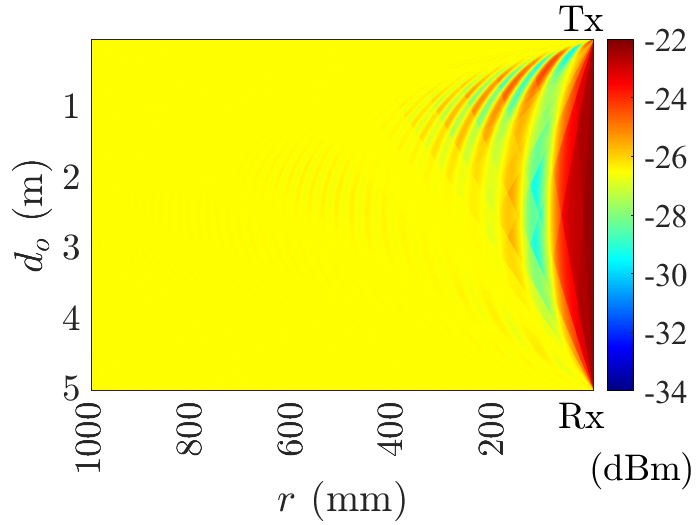}
	\end{minipage}}
\caption{The field of view of main and guard beams}
\label{fig:simulated_fov_guard_beam}
\end{figure}
\vspace{-1mm}

\subsection{Rx field of view}

Fig.~\ref{fig:simulated_fov_guard_beam} shows the Rx field of view of the main beams and guard beams, describing the Fresnel zone around Tx and Rx that are separated at $d_o$ = 5~m. The field of view is represented as the combined received signal level of main beam and guard beam $z(p)$ when the blocker is at point $p$ within the detection area. 
For this, only a single reflection point is considered. The points where the signal experiences constructive and destructive interference are depicted by the curve patterns with different colours, indicating the received signal level changes when the blocker approaches the Tx and Rx. This explains why the received signal fluctuates heavily during the pre-shadowing event.

When only using a main beam with $\Theta_R$~=~7\textdegree, the curved area is symmetrical because Tx has the same beam width ($\Theta_T$~=~7\textdegree) as the Rx. 
As described in Fig.~\ref{fig:fov_main_7}, almost no curve pattern is observed beyond $r$ = 200 mm due to highly directional beams used in Tx and Rx. Consequently, the presence of a blocker can only be detected when it is very close to the Tx-Rx direct path. On the other hand, by using additional guard beam (i.e. $\Theta_R'$~=~7\textdegree, $\Phi$~=~7\textdegree~and $\Theta_R'$~=~7\textdegree, $\Phi$~=~14\textdegree), the curved patterns are observed within a broader range $r$, as shown in Fig.~\ref{fig:fov_guard_7_7} and \ref{fig:fov_guard_7_14}, respectively. Thus, the fluctuation can be detected earlier when the blocker is still further away from Tx-Rx direct path compared to when only using the main beam.

\subsection{The detection range}

\begin{figure}[t]
    \centering
    \includegraphics[width=0.9\linewidth]{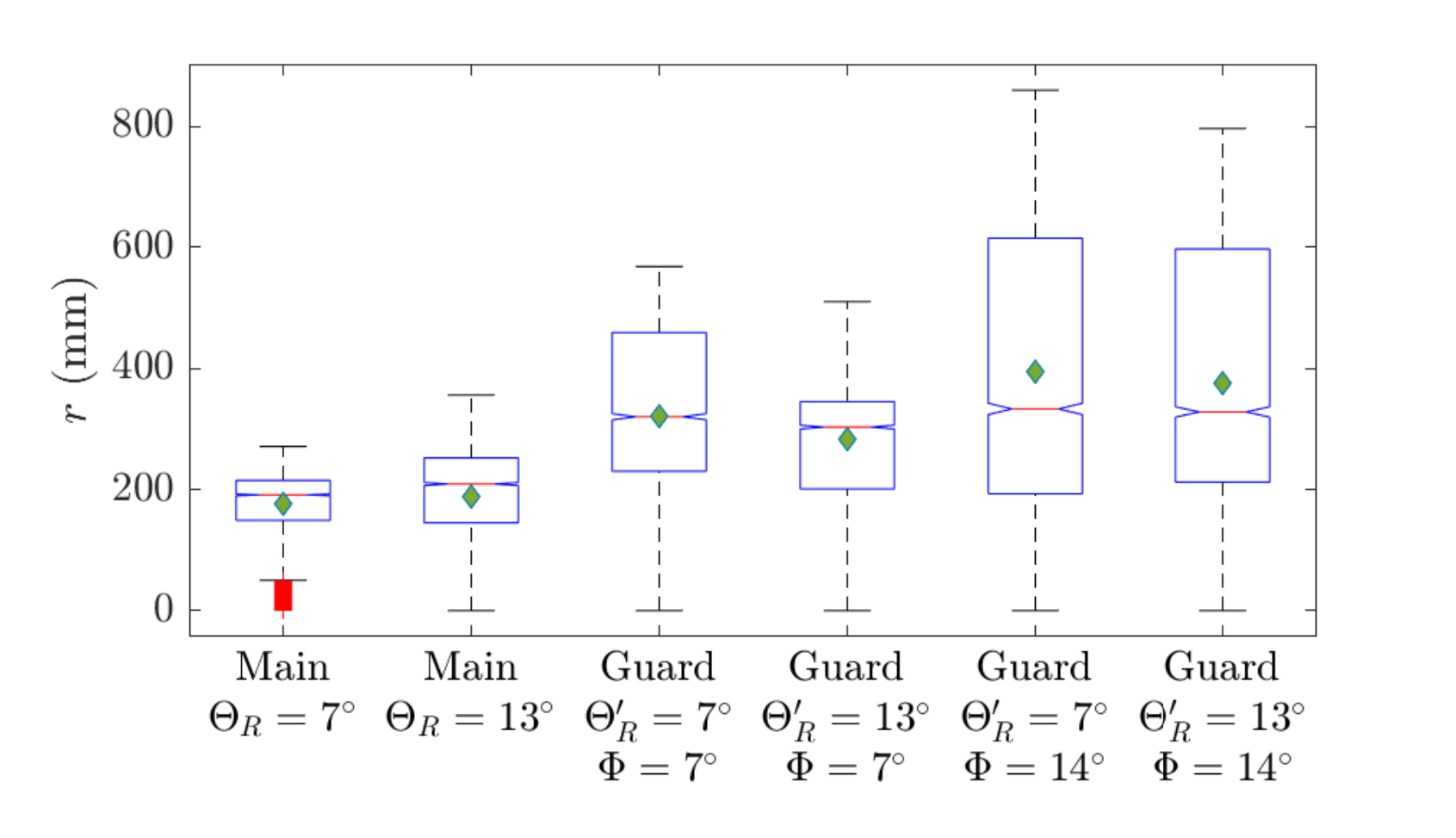}
    \caption{\centering{Detection range of various Rx beam configurations}} 
    \label{fig:detection_range}
\end{figure}
\vspace{-1mm}

The detection range comparison of main and guard beams with various Rx beam configurations is shown in Fig.~\ref{fig:detection_range}, where the green diamond indicates the average detection range of each configuration. 
Increasing the beamwidth of the main beam from $\Theta_R$~=~7\textdegree ~to $\Theta_R$~=~13\textdegree ~improves the detection range slightly. However, this improvement comes with the cost of lowering the channel gain as depicted in Fig.~\ref{fig:fov_main_13}, leading to a reduced communication performance. By leveraging a guard beam, the performance of the main communication beam will not be affected. 

At steering angle $\Phi$ = 7\textdegree, the guard beam with $\Theta_R'$~=~7\textdegree ~provides $\sim$140~mm better average detection range compared to the main beam only with $\Theta_R$~=~7\textdegree. 
By increasing the steering angle to $\Phi$ = 14\textdegree, the average detection range improves significantly where the achievable detection range is up to $\sim$860~mm. 
The guard beam with wider beamwidth ($\Theta_R'$~=~13\textdegree) does not provide better detection range compared to the narrower one. Although the field of view is expanded, weaker level of NLOS components are received due to lower beamforming gain, thus causing less Rx level fluctuation in this configuration.

\section{Experimental Evaluation}
\label{sec:measurements}

The blockage prediction time by main and guard beams, as well as the prediction accuracy, are evaluated through measurement campaigns using our mmWave MIMO testbed. 

\subsection{Measurement setup and scenario}

\begin{figure}[t]%[tbh]
\centering
\begin{minipage}{.235\textwidth}
  \centering
  \includegraphics[width=1.0\linewidth]{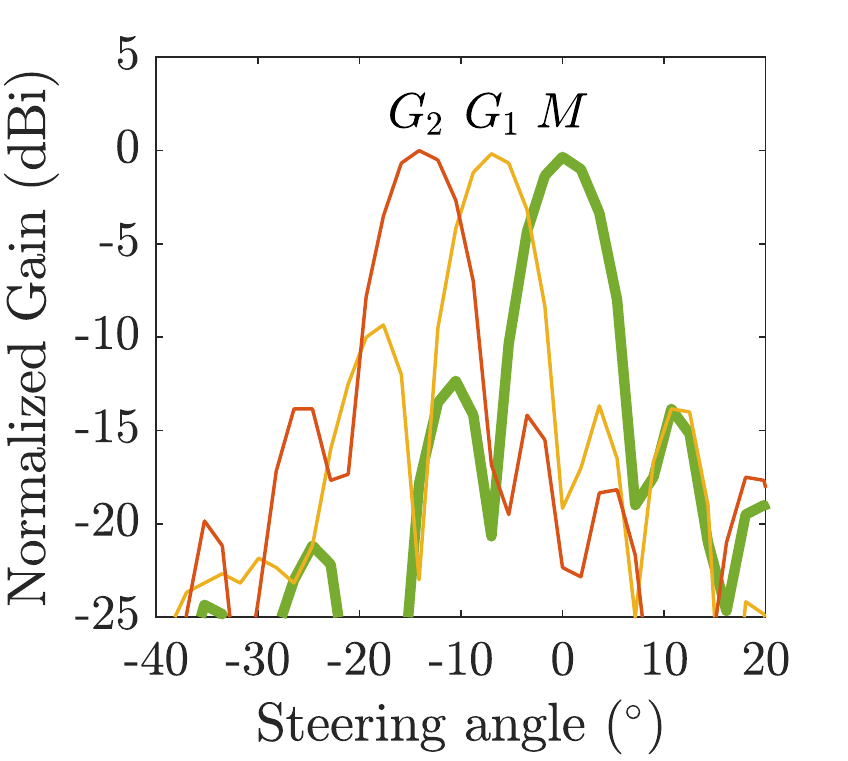}
  \captionof{figure}{The main and guard beam patterns}
  \label{fig:beam_pattern}
\end{minipage}%
\begin{minipage}{.235\textwidth}
  \centering
  \includegraphics[width=1\linewidth]{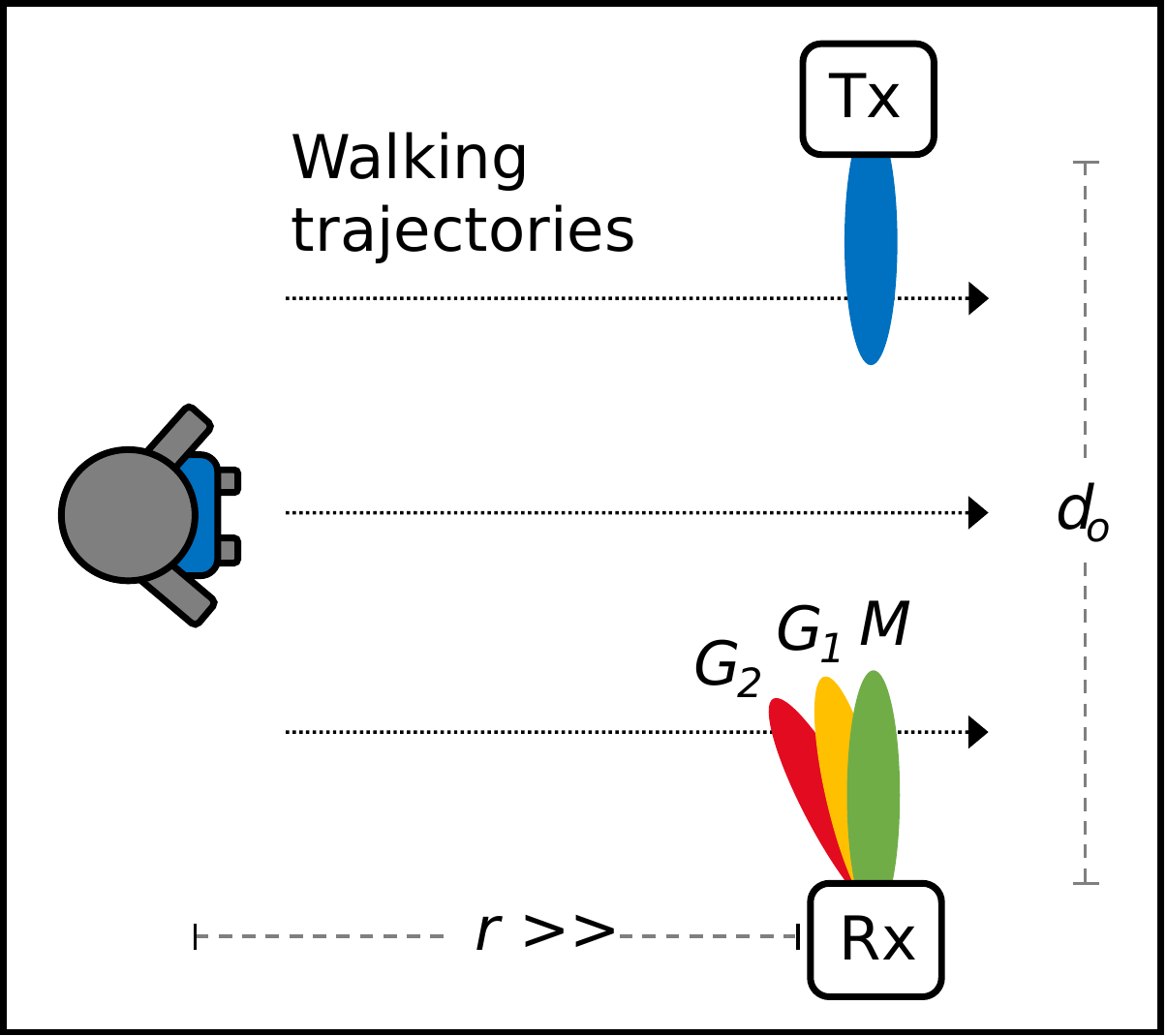}
  \captionof{figure}{Measurement scenario}
  \label{fig:measurement_setup}
\end{minipage}
\end{figure}
\vspace{-1mm}

\label{subsec:hardware}
We use Butler Matrix units and MIMO testbed consisting of multiple USRPs for the measurement. The Butler Matrix is a mmWave front-end operating in the 25-30 GHz frequency range. A Butler Matrix unit can either transmit or receive up to 16 orthogonal beams in which the beam patterns depend on the frequency used~\cite{Xiaozhou2019}.

A Tx USRP transmitting the intermediate frequency $f_{IF}$~=~2.4~GHz is used to generate a mmWave signal at $f$~=~26.0~GHz. The $f_{IF}$ is fed to one of the Tx Butler Matrix input ports while an RF signal generator operating at $f_{LO}$~=~11.8~GHz is connected to the Local Oscillator (LO) port of the Butler Matrix to upconvert $f_{IF}$ to $f$. The up/down-conversion takes place in a Butler Matrix unit with the following calculation: ${f = 2 f_{LO} + f_{IF}}$. The generated Tx beam has HPBW $\Theta_T$~=~7\textdegree ~\cite{Colpaert2020}.

On the receiver side, three orthogonal beams are chosen to represent the main beam and guard beams with different steering angles $\Phi$ as shown in Fig.~\ref{fig:beam_pattern}. The main beam ($M$) has $\Theta_R$~=~7\textdegree ~and is aligned with the direction of the Tx beam. The first guard beam ($G_1$) has $\Theta_{R1}'$~=~7\textdegree ~\& $\Phi_{1}$~=~7\textdegree, and the second guard beam ($G_2$) has $\Theta_{R2}'$~=~7\textdegree ~\& $\Phi_{2}$~=~14\textdegree. 

Both Tx and Rx USRPs run LabView Communications MIMO Application Framework\cite{ni_mimo} in which Tx and Rx USRPs are synchronized using Pulse Per Second and 10 MHz reference signals. The framework uses a Time Division Duplex signal frame structure with OFDM symbols in which the channel estimation is based on the uplink pilot. The list of measurement parameters is presented in Table~\ref{tab:parameters}.

The measurements are conducted in an indoor environment with a room size of 11$\times$6.5 m{$^{2}$}. The setup consists of a pair of Tx and Rx Butler Matrix separated in $d_o$~=~5~m with the height of Tx and Rx being 1 m from the ground.  
The main and guard beams capture the complex channel state information during 5~s to give enough time for a person to cross the Tx-Rx link. During a measurement run, a person walks in the direction of one out of three possible trajectories, as depicted in Fig.~\ref{fig:measurement_setup}. In total, 300 blockage samples are recorded. In each measurement sample, 500 time symbols are captured every 10 ms, in which each symbol contains 100 complex channel state information samples of 100 subcarriers for each receive beam.  

\begin{figure}[t]
  \subfloat[Rx level fluctuation]{
	\begin{minipage}[b]{0.2495\textwidth} \label{fig:example_rx_fluctuation}
	   \centering
	   \includegraphics[width=1.0\textwidth]{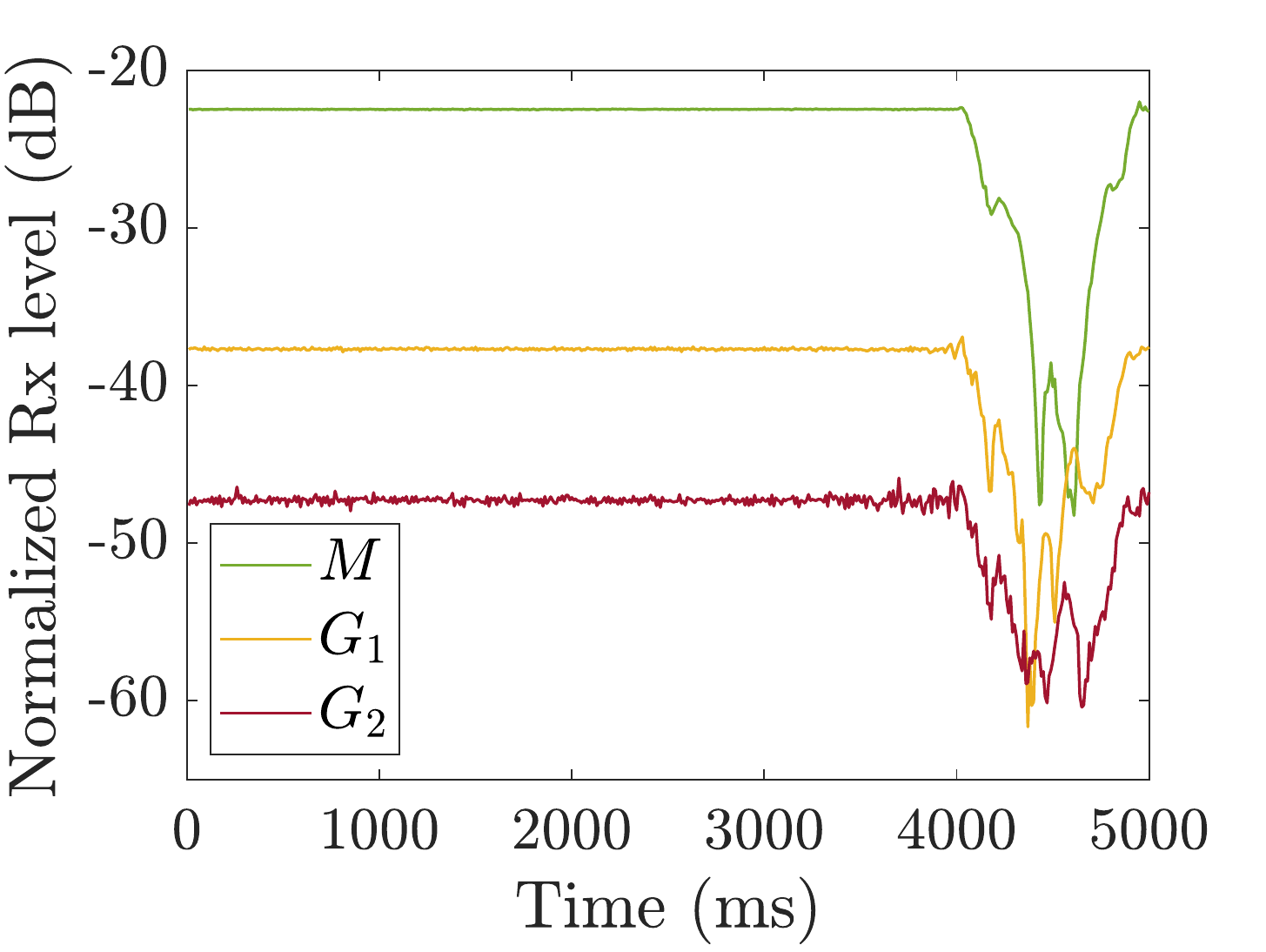}
	\end{minipage}}%
	\subfloat[Rx level standard deviation]{
	\begin{minipage}[b]{0.2495\textwidth} \label{fig:example_std_rx}
	   \centering
	   \includegraphics[width=1.0\textwidth]{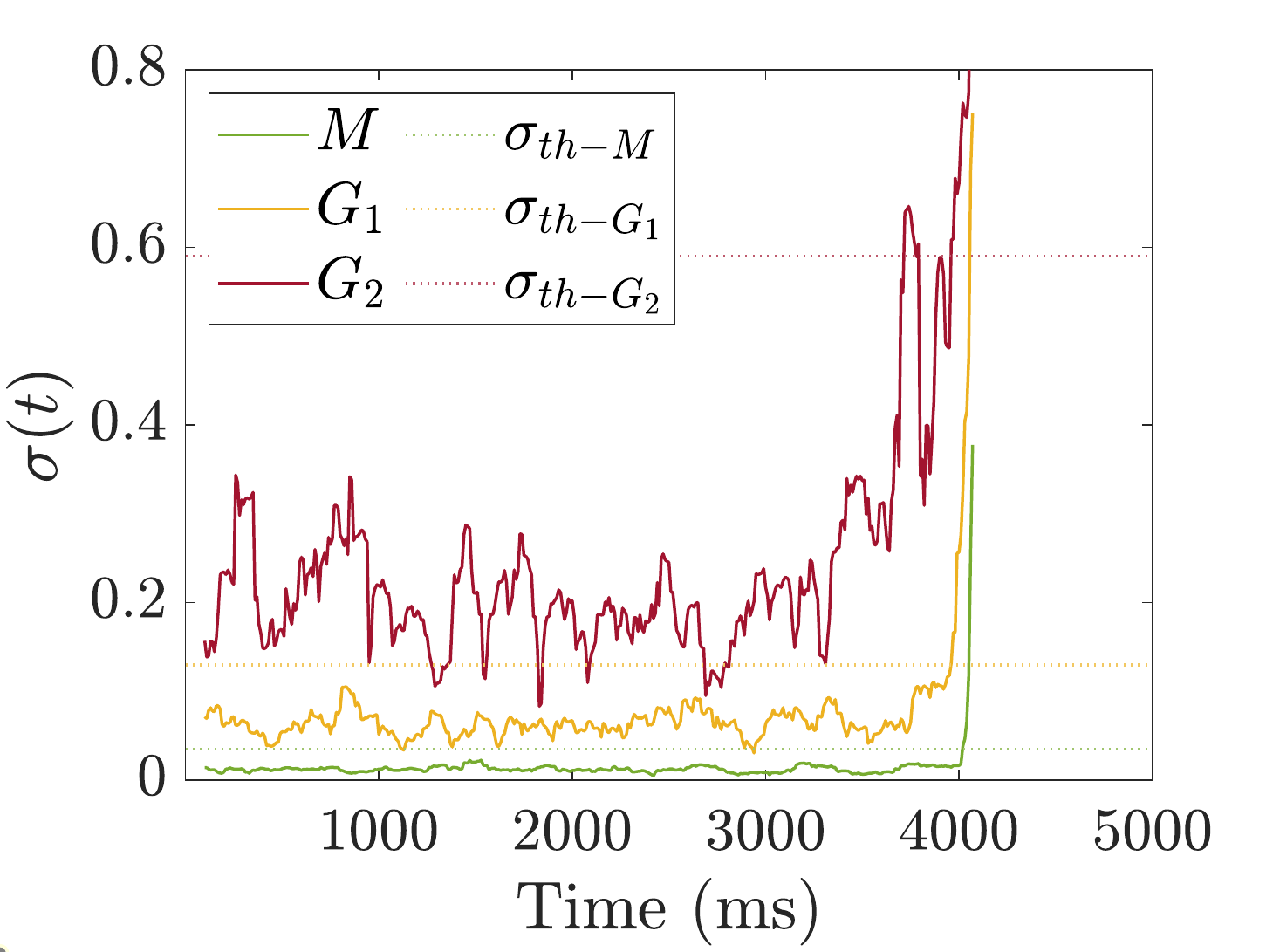}
	\end{minipage}}
\caption{Example of pre-shadowing event}
\end{figure}
\vspace{-1mm}

\subsection{Blockage prediction time}

\begin{figure}[t]
\centering
\begin{minipage}{.248\textwidth}
  \centering
  \includegraphics[width=1.0\linewidth]{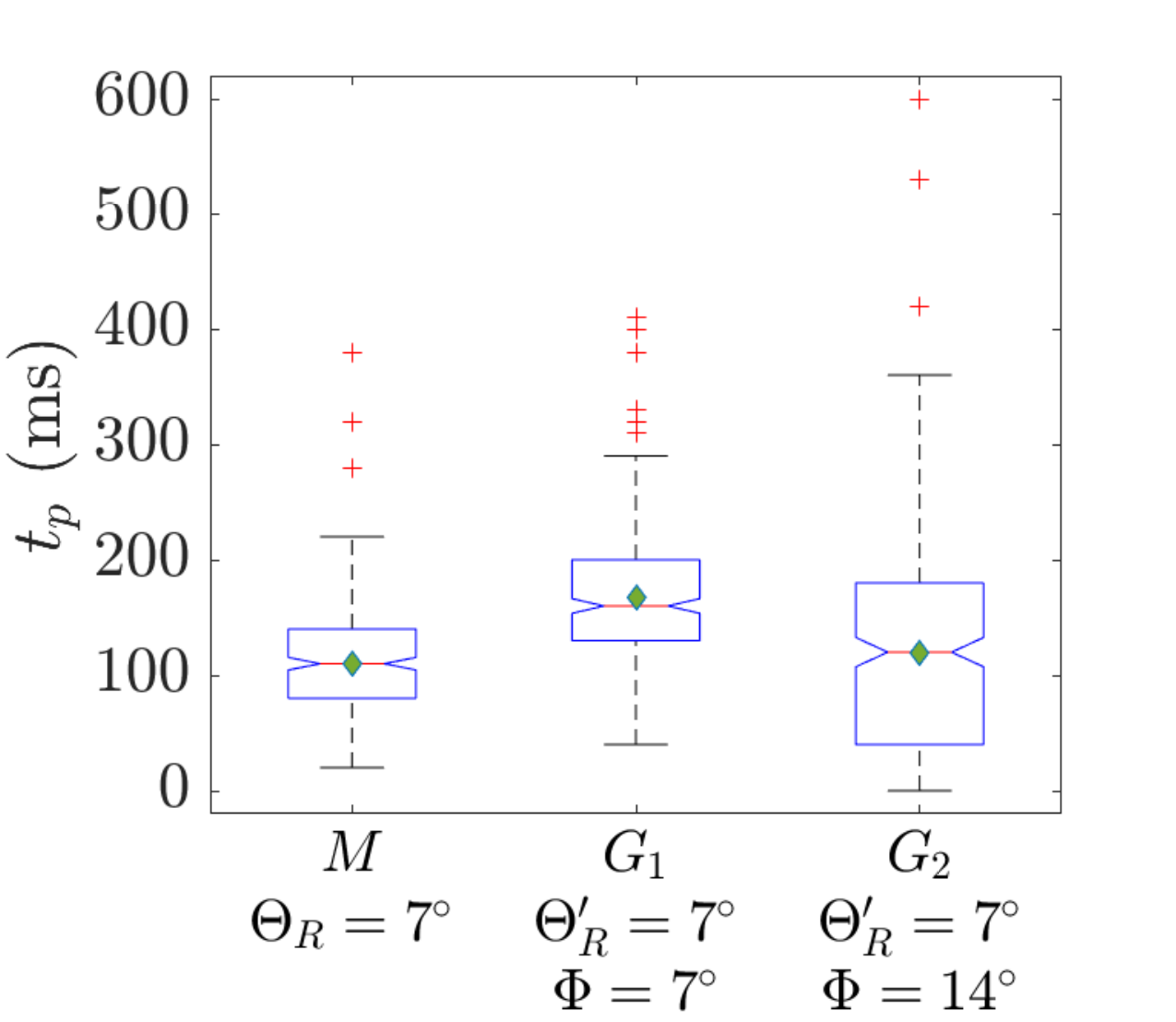}
  \captionof{figure}{Measurement-based prediction time}
  \label{fig:measurement_results}
\end{minipage}%
\begin{minipage}{.248\textwidth}
  \centering
  \includegraphics[width=1.0\linewidth]{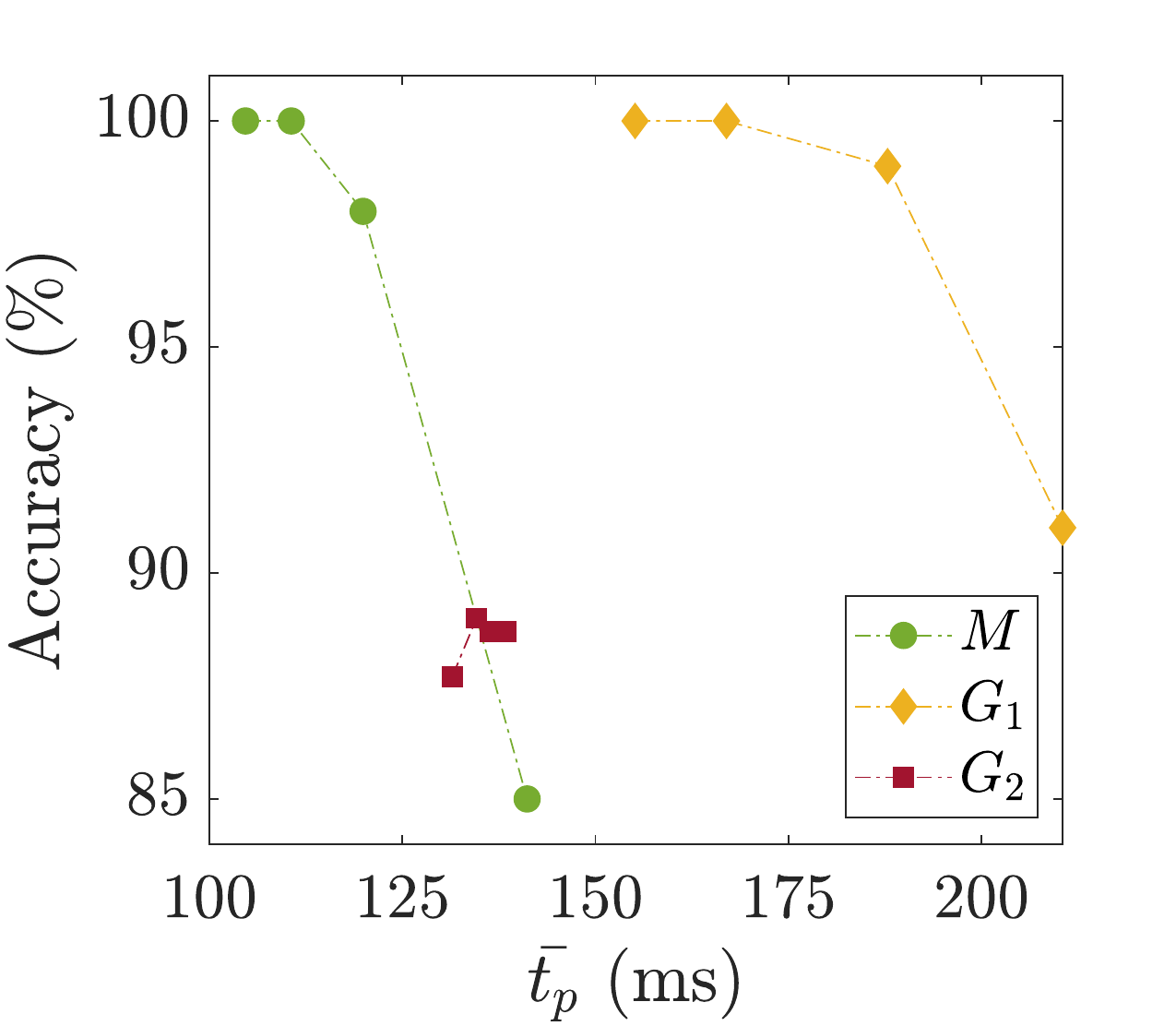}
  \captionof{figure}{Prediction time vs. accuracy}
  \label{fig:tradeoff_time_vs_accuracy}
\end{minipage}
\end{figure}
\vspace{-1mm}

The blockage prediction time ($t_p$) metric indicates how early the human blockage can be detected depending on the Rx signal volatility. As shown in Fig.~\ref{fig:example_rx_fluctuation}, different Rx beam configurations fluctuates differently. The main beam $M$ has the highest Rx level and the most stable one, followed by guard beams $G_1$ and $G_2$. Due to the different levels of standard deviation, different $\sigma_{th}$ values are applied to all compared three Rx beam configurations as indicated in Table~\ref{tab:parameters}.   

Fig.~\ref{fig:measurement_results} shows the prediction time distribution of three different Rx beam configurations in which the green diamond indicates the average prediction time $\bar{t_p}$. For the main beam $M$, the achievable average prediction time is 110.6~ms, comparable with the achievable detection time in \cite{Koda2020}.
$G_1$ and $G_2$ provide $\bar{t_p}$ of 166.97~ms and 119.8~ms, with the achievable ${t_p}$ of up to 290~ms and 360~ms, respectively. 

On average, $G_1$ provides 1.5 times earlier $\bar{t_p}$ than $M$ while $G_2$ does not provide a significant $\bar{t_p}$ improvement over $M$ since the false detection and misdetection in $G_2$ occur more frequently compared to two other configurations. A larger steering angle of $G_2$ is intended to detect the blocker in a further range. However, since the power transmitted to illuminate the blocker in that range is small due to a narrow Tx beam, a low reflected signal level is received by $G_2$. It makes the channel become highly affected by the noise.
Consequently, $\sigma(t)$ fluctuates highly even without the presence of a blocker. The instability of $\sigma(t)$ leads to a false detection depending on the $\tau_d$ and $\sigma_{th}$ values. 
Results in Fig.~\ref{fig:measurement_results} represent the $t_p$ distribution when the achievable prediction accuracy of $G_2$ is 89\%, while both $M$ and $G_1$ obtain 100\% accuracy.

\subsection{Trade-off: prediction time vs. prediction accuracy}

Increasing $\sigma_{th}$ may reduce the false detection, thus improving the accuracy at the cost of $\bar{t_p}$ reduction. The trade-off between the prediction time and accuracy is described in Fig.~\ref{fig:tradeoff_time_vs_accuracy}. Different $\sigma_{th}$ values are applied to evaluate the prediction time and the accuracy of those compared Rx beam configurations.  

For $M$, lowering $\sigma_{th}$ to 0.025 increases $\bar{t_p}$ to 141~ms at the cost of accuracy reduction to nearly 85\%. For $G_1$, $\bar{t_p}$ of 210~ms with 91\% accuracy can be achieved by lowering the $\sigma_{th}$ to 0.11. Increasing $\sigma_{th}$ for $M$ and $G_1$ to 0.04 and 0.14, respectively, will only reduce the $\bar{t_p}$ without affecting the accuracy. On the other hand, increasing $\sigma_{th}$ to 0.59 in $G_2$ reduces its accuracy due to misdetection. 
In practice, $\sigma_{th}$ must be appropriately adjusted depending on the environment to achieve an optimal prediction time with high accuracy.  

\section{Conclusion}
\label{sec:conclusion}

This paper proposes an early predictive method for anticipating a dynamic blockage in mmWave communication by leveraging a guard beam, an additional passive Rx beam next to the primary communication beam. It aims to expand the detection range so that the blocker approaching the communication beam can be detected earlier through an in-band Rx level volatility observation during the pre-shadowing event. Using a guard beam prevents the communication beam performance from being sacrificed for better blockage detection without requiring any additional detection system.  
Pre-shadowing event channel models for both main and guard beams are proposed. Our evaluation shows that the guard beam can extend the detection range up to 860~mm and predict the blockage up to 360~ms before the shadowing occurs, benefiting early dynamic blockage detection. 
For future work, we will evaluate the prediction accuracy with the presence of multiple moving blockers walking in different trajectories and velocities.

\section*{Acknowledgment}
This work has received funding from the EU-H2020 MSCA-ITN-2019 MINTS project No. 861222 and the MSCA-IF-2020 V.I.P. project No. 101026885.

\vspace{-1.2mm}
\bibliographystyle{IEEEtran}
\bibliography{./mybib}

\end{document}